\documentclass[runningheads]{llncs}
\usepackage{graphicx}
\usepackage{xcolor}
\usepackage[short]{optidef}
\usepackage{amssymb}
\usepackage{siunitx}
\usepackage{booktabs}
\usepackage{multirow}
\begin{document}
\title{Influence of Incremental Constraints on Energy Consumption and Static Scheduling Time for Moldable Tasks with Deadline} 
%
\titlerunning{Incremental Constraints for Static Scheduling}
%
\author{J\"org Keller\orcidID{0000-0003-0303-6140} \and
Sebastian Litzinger\orcidID{0000-0003-2200-7337} 
}
\authorrunning{J. Keller et al.}
%
\institute{FernUniversit\"at in Hagen, Hagen, Germany\\
\email{firstname.lastname@fernuni-hagen.de}}
\maketitle              
\begin{abstract}
Static scheduling of independent, moldable tasks on parallel machines with frequency scaling comprises decisions on core allocation, assignment, frequency scaling and ordering, to meet a deadline and minimize energy consumption.
Constraining some of these decisions reduces the solution space, i.\,e. 
may increase energy consumption, but may also
reduce scheduling time or give the chance to tackle larger task sets.  
We investigate the influence of different constraints that lead from an unrestricted scheduler via two intermediate steps to the crown scheduler, by presenting integer linear programs for all four schedulers.
We compare scheduling time and energy consumption for a benchmark suite of synthetic task sets of different sizes.
Our results indicate that the final step towards the crown scheduler -- the execution order constraint -- is responsible for faster scheduling when task sets are small, and lower energy consumption when we deal with large task sets.

\end{abstract}
\section{Introduction} \label{sec:intro}

Parallel programs can often be formulated as a set of tasks working on a stream of inputs \cite{MelotKEK19}. While the tasks may exhibit dependencies, task instances working on different inputs are independent. Throughput requirements impose a maximum deadline until all task instances must be executed, while the 
nature of the platform
--- often embedded or mobile --- 
necessitates to restrict energy consumption as much as possible. To meet the deadline, it is often necessary to parallelize some or all tasks.

Static scheduling of independent, parallelizable tasks of known workloads with a common deadline onto a parallel machine with frequency-scalable cores comprises a
 number of steps.
For moldable tasks, i.\,e. tasks where the degree of parallelization must be fixed prior to execution and cannot be changed during execution, these steps are allocation, mapping, scaling and ordering. This means that for each task, the degree of parallelism must be determined and an appropriate subset of cores must be assigned, the operating frequency chosen from the available frequency levels and the tasks ordered in time so that they do not overlap.
The settings must be such that all tasks terminate before the common deadline.
Among the schedules that achieve this, it is often desirable to choose one that minimizes another property such as the energy consumption by the tasks, which is mainly determined by the tasks' execution frequencies.

The decisions in these phases are not independent of each other. Especially, any sub-optimal decision in the allocation, mapping or ordering phases that would result in missing the deadline can only be 
compensated
by increasing the operating frequencies (assuming that maximum frequency will always suffice to meet the deadline), which in turn increases the energy consumption.

\begin{figure}
    \centering
    \includegraphics[width=0.47\textwidth]{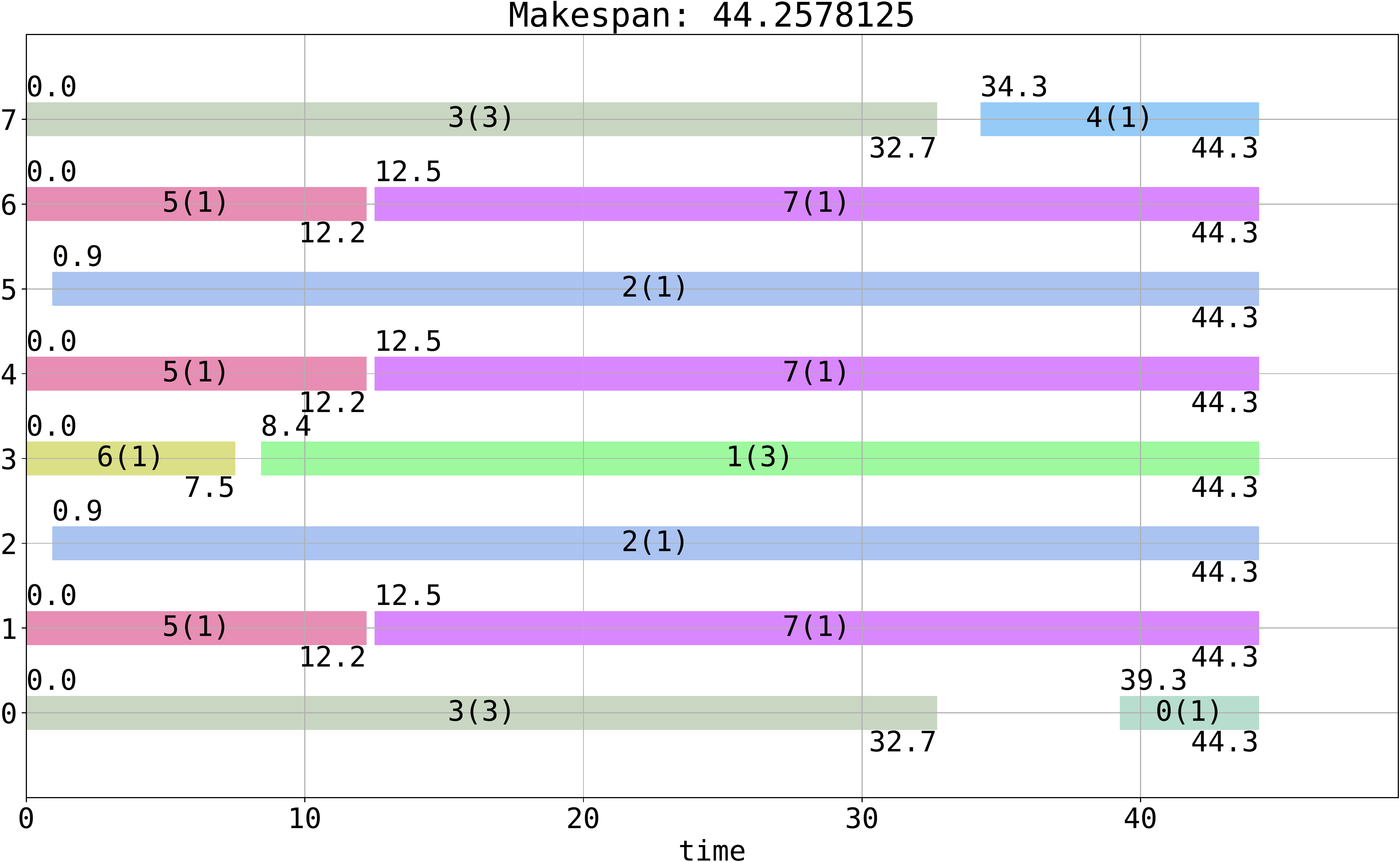}\qquad \includegraphics[width=0.47\textwidth]{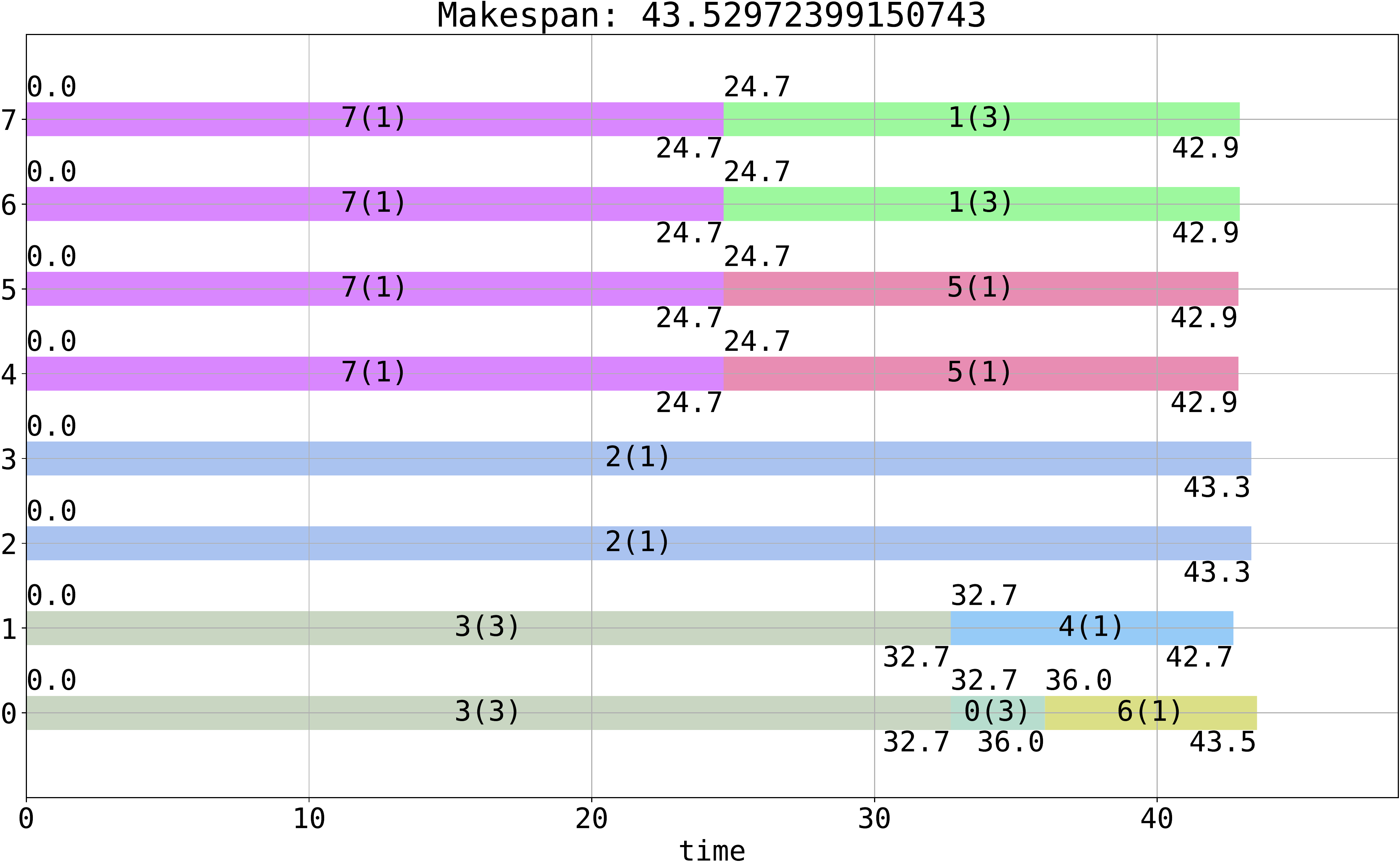}
    \caption{Example schedules for moldable tasks: (a) unrestricted schedule (left), (b) crown schedule (right).}
    \label{fig:exschedules}
\end{figure}

An unrestricted schedule with the above properties looks like the solution of a kind of puzzle game (cf.~Fig.~\ref{fig:exschedules}(a)). It can be computed by an integer linear program (ILP) \cite{MelotKEK19}, which however needs a large number of variables and thus can only be used for very moderate task counts.

\emph{Crown Scheduling} \cite{MelotKKE14} is an approach where allocation, mapping and ordering are 
restricted (cf.~Fig.~\ref{fig:exschedules}(b)) in order to reduce the number of decision variables in the ILP to allow scheduling of larger task sets or reduce scheduling time for the same task set compared to an unrestricted scheduler.
The price to pay is that the solution space for crown scheduling is only a subset of the solution space of the unrestricted scheduler, so that the crown-optimal solution found may have a higher energy consumption.

It turns out \cite{MelotKEK19} that crown scheduling only leads to a moderate increase in energy consumption in most cases but often 
reduces the scheduling time and/or the number of ILP solver timeouts compared to an unrestricted scheduler.

In the present research, we investigate the influence which each of the above steps has 
on energy consumption and scheduler execution time when crown scheduling constraints are applied. Thus, we compare four schedulers:
\begin{itemize}
    \item unrestricted scheduler,
    \item unrestricted scheduler with allocation constrained to powers of two,
    \item unrestricted scheduler with constrained allocation and assignment 
    restricted
    to consecutive cores starting with a core whose index is a multiple of the allocation (e.\,g., for an allocation of 4, possible assignments are $\{0,1,2,3\}$, $\{4,5,6,7\}$, and so on),
    \item and crown scheduling, where, additionally, the tasks are ordered in time such that tasks with larger allocation are always executed before tasks with smaller allocation.
\end{itemize}

We compare the schedulers with a benchmark of synthetic task sets of different sizes and evaluate both energy consumption and scheduling time of the different schedulers.
We find that the last of the above mentioned steps is decisive in either reducing scheduling time or obtaining a higher quality solution if the solver's timeout is reached, depending on task set size.

The remainder of the article is structured as follows. 
In Section~\ref{sec:background}, we present background information on static scheduling and energy efficiency, and discuss related work.
Section~\ref{sec:constraints} presents integer linear programs for static schedulers 
implementing progressively more expansive restrictions, 
starting from an unrestricted scheduler and arriving at a crown scheduler.
In Section~\ref{sec:eval}, we report on experiments with synthetic task sets to compare the different schedulers, 
while Section~\ref{sec:concl} concludes and gives an outlook onto future work.

\section{Background}\label{sec:background}

\subsection{Task Scheduling}

We assume that we have $n$ independent tasks $\tau_j$, where $j=1,\ldots,n$, each with its workload $\lambda_j$, given in number of cycles. Thus, if the task is run at frequency $f$, its runtime is $\lambda_j/f$.
Each task can be parallelized on up to $W_j$ cores. Since we deal with moldable tasks, the number of cores $w_j$ it runs on cannot change during execution, and non-preemption prohibits suspension and subsequent continuation of a task. Moreover, a task-individual function $e_j(q)$ specifies its parallel efficiency when run on $q$ cores. A task is executed at one of $K$ operating frequencies $f_k$, $k = 0,\ldots,K-1$, its per-core runtime can then be determined as $t_j(w_j,f_k) = \lambda_j / (f_k \cdot w_j \cdot e_j(w_j))$. All tasks shall be completed until a given deadline $M$.

We furthermore assume that the task set is to be scheduled to a homogeneous machine with $p$ cores $P_0,\ldots,P_{p-1}$, where each core can be scaled to one of the available discrete operating frequencies $f_k$ independently of the other cores. In the following, we take the frequency scaling overhead to be negligible. For all frequency levels, the corresponding per-core power consumption $Pow(f_k)$ is known and assumed to be constant.\footnote{Of course, core power consumption does not only vary with the operating frequency. Since the instruction mix executed by the processor affects power consumption as well \cite{seo:2015}, 
it can also be task (type)-specific, which we do not model here. Other factors influencing a core's power consumption are the voltage, which we assume is always set to the least possible value for the given frequency, and the chip temperature, which can be kept in check via cooling.}

Scheduling a task set to a given machine then consists of a number of steps, which may be performed subsequently one at a time, partially conjoined, or even all at the same time.
By \emph{allocation} we understand that for each task, the number of cores it should run on is determined.
By \emph{mapping} we understand that the task is assigned to a subset of cores for execution, where the size of this subset must correspond to the task's allocation.
By \emph{scaling} we understand that the task is assigned an operating frequency to determine its runtime.
By \emph{ordering} we understand that each task is assigned a start time (and thus an end time, as the runtime is known), such that no two tasks overlap in execution, i.\,e. if two tasks' assignments are not disjunct, the tasks are not allowed to overlap in time (so-called \emph{feasible} schedule, for examples of a feasible schedule see e.\,g. Figure \ref{fig:exschedules}) and must be ordered such that the start time of one task must be at least the end time of the other task or vice versa.

\subsection{Energy Consumption}

The energy required for executing a task $\tau_j$ can be computed as the product of its per-core runtime, the core's power consumption at the designated operating frequency $f_k$, and the number of cores the task runs on: 
\begin{gather*}
E_j = t_j(w_j,f_k) \cdot Pow(f_k) \cdot w_j.
\end{gather*}
The total energy consumption the execution of a schedule causes is the sum of all the tasks' energy consumption values:
\begin{gather*}
E_{\text{total}} = \sum_j E_j.
\end{gather*}
Here, we do not model the energy consumption when cores are idle, as we choose the deadlines sufficiently tight for long idle periods not to occur.

When scheduling under a deadline constraint one can choose among all feasible schedules, i.\,e. schedules not violating the deadline (as long as there is more than one). This creates the potential to optimize for some other feature, which guides said choice accordingly. In the current paper, we opt for minimizing the energy consumption during the schedule's execution, see
Section \ref{sec:constraints}. 

\subsection{Related Work}


Most research in the area of scheduling consider to either find optimal solutions or a particular approach to constrain the large solution space.
Turek \textit{et al.}~\cite{Turek1992} consider scheduling of moldable tasks on multiprocessors with the goal of makespan minimization and give approximations.
Pruhs \textit{et al.}~\cite{pruhs2008speed} present an optimal scheme, but they only consider sequential tasks, assume continuous frequencies, and optimize makespan for a given energy budget.
Sanders and Speck~\cite{SandersSpeck2012} investigate energy-efficient scheduling for malleable tasks with preemption, while we consider moldable tasks and non-preemption.
Zahaf \textit{et al.}~\cite{Zahaf2017} present a solution to schedule moldable tasks, but their solution uses non-linear integer programming, and their focus is on heterogeneity of the platform and on modelling of the power consumption.
Xu \textit{et al.}~\cite{xu12} propose optimal and heuristic solutions
to schedule moldable tasks. They use a bookshelf approach to order tasks, which however seems inferior to crown scheduling \cite{MelotKKE14}.
Crown scheduling \cite{MelotKKE14} applies a particular set of constraints, but only compares to other constrained and unrestricted \cite{MelotKEK19} schedulers.
Ye \textit{et al.}~\cite{Ye2018} investigate online scheduling of moldable task sets to minimize makespan, while we consider static scheduling to minimize energy under a deadline constraint.

\section{Schedulers with different Constraints}\label{sec:constraints}

The most basic scheduler, which marks our starting point, is the unrestricted scheduler. The constraints applying here solely ensure the resulting schedule's feasibility but do not impose any further limitations. To compute a schedule, the scheduler solves an ILP with $p \cdot n \cdot K$ decision variables $x_{i,j,k}$, another $p \cdot n \cdot K$ decision variables $z_{i,j,k}$, $n^2$ decision variables $y_{j,j'}$, $n$ decision variables $s_j$, and $n$ decision variables $e_j$. The underlying semantics is as follows:
\begin{itemize}
\item $x_{i,j,k} = 1$ iff $\tau_j$ runs on $i$ cores on frequency level $k$,
\item $z_{i,j,k} = 1$ iff $\tau_j$ runs on $P_i$ at frequency $f_k$,
\item $y_{j,j'} = 1$ iff $\tau_j$ precedes $\tau_{j'}$ on one or more cores $\tau_{j'}$ runs on,
\item $s_j$ is the time when execution of $\tau_j$ commences,
\item $e_j$ is the time when execution of $\tau_j$ terminates.
\end{itemize}
As discussed in Section \ref{sec:background}, the corresponding ILP minimizes the energy required for executing the resulting schedule:

\begin{mini!}[2]
{}
{E_{\text{total}} = \sum_{i,j,k} x_{i,j,k} \cdot t_j(w_j,f_k) \cdot Pow(f_k) \cdot w_j}
{}
{}
\addConstraint{\forall j \quad \sum_{i,k} x_{i,j,k}}{=1\label{constr.b}}
\addConstraint{\forall j \quad e_j}{\leq M\label{constr.c}}
\addConstraint{\forall j \quad s_j}{\geq 0\label{constr.d}}
\addConstraint{\forall j \quad e_j}{=s_j + \sum_{i,k} x_{i,j,k} \cdot t_j(w_j,f_k)\label{constr.e}}
\addConstraint{\forall j \quad y_{j,j}}{=0\label{constr.f}}
\addConstraint{\forall j,j' \quad y_{j,j'} + y_{j',j}}{\leq 1\label{constr.g}}
\addConstraint{\forall j,j' \neq j \quad s_j}{\geq e_{j'} - (1 - y_{j',j}) \cdot M\label{constr.h}}
\addConstraint{\forall j,j'<j,i \quad y_{j,j'} + y_{j',j}}{\geq \sum_{k} z_{i,j,k} + z_{i,j',k} - 1\label{constr.i}}
\addConstraint{\forall j,k \quad \sum_i z_{i,j,k}}{=\sum_i i \cdot x_{i,j,k}.\label{constr.j}}
\end{mini!}

As with all the schedulers presented in this section, the objective function to be minimized is the total energy consumption $E_{\text{total}}$. Constraint \eqref{constr.b} ensures that each task is scheduled exactly once. Constraints \eqref{constr.c} and \eqref{constr.d} guarantee that each task starts and completes execution in $[0,M]$, while \eqref{constr.e} ties $e_j$ to $s_j$ by setting $e_j$ to the sum of $s_j$ and $\tau_j$'s per-core runtime. Constraint \eqref{constr.f} prohibits self-precedence, and \eqref{constr.g} mutual precedence of any two tasks. Constraint \eqref{constr.h} ensures that a task's execution can only begin if all preceding tasks have completed. Constraint \eqref{constr.i} forces specifying a preference relation for tasks sharing one or more cores. Finally, \eqref{constr.j} ascertains consistency of allocation and mapping for each task.

Moving from the unrestricted scheduler to the allocation-constrained schedules requires the introduction of an additional constraint:
\begin{gather*}
\forall j,i : \log_2(i) \notin \mathbb{N} \quad \sum_k x_{i,j,k} = 0.
\end{gather*}
Thus, allocations which are not powers of 2 are banned.

\begin{figure}
    \centering
    \includegraphics[width=0.8\textwidth]{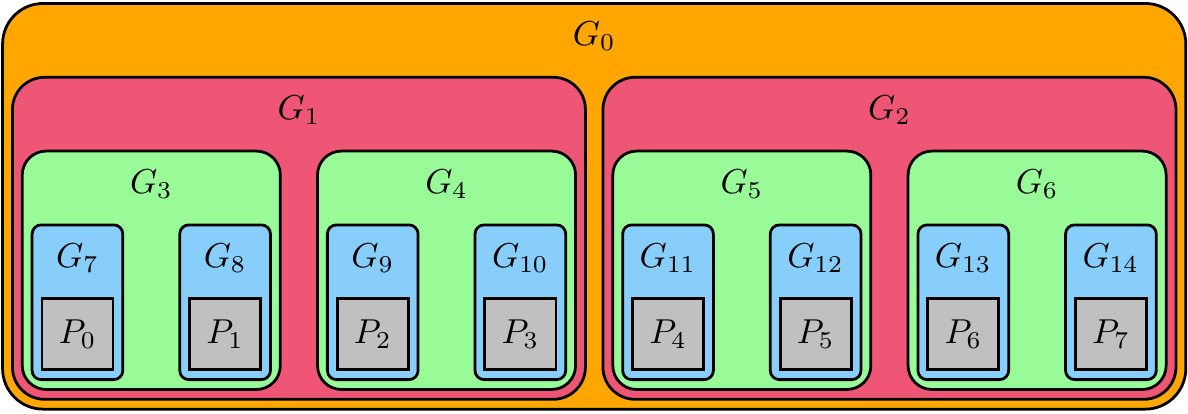}
    \caption{Core group structure of a processor with 8 cores}
    \label{fig:coregroups}
\end{figure}

Proceeding to the group scheduler, we establish the concept of core groups as in \cite{MelotKKE14}, cf. Figure \ref{fig:coregroups}. We now have $2p-1$ core groups of different sizes. The \emph{root} group $G_0$ comprises all cores. It is decomposed into the disjoint and equally sized groups $G_1$ (ranging over $P_0$ to $P_3$) and $G_2$ (spanning $P_4$ to $P_7$), which are in turn divided in the same fashion, and so on. Ultimately, the \emph{leaf} groups ($G_7$ to $G_{14}$ in Figure \ref{fig:coregroups}) contain one core only.

The decision variables $z_{i,j,k}$ are not needed for the group scheduler and therefore are dropped from the ILP. For the decision variables $x_{i,j,k}$, the semantics must be modified as follows:
\begin{itemize}
    \item $x_{i,j,k} = 1$ iff $\tau_j$ runs in core group $G_i$ at frequency $f_k$.
\end{itemize}
We furthermore acknowledge that $w_j = p_i$ when $\tau_j$ is mapped to $G_i$, $p_i$ being the number of cores in $G_i$. While constraints \eqref{constr.b} to \eqref{constr.h} remain as they are, \eqref{constr.j} is removed, and \eqref{constr.i} is replaced by the following restriction:
\begin{gather*}
\forall j,j'<j,i \quad y_{j,j'} + y_{j',j} \geq \sum_{k} x_{i,j,k} + \sum_{g \in \mathit{offspring}(G_i),k} x_{g,j',k} - 1.
\end{gather*}
That way, we make sure that a preference relation holds whenever two tasks are in the same group or one is in a subgroup of the other's group. Here, $\mathit{offspring(G_i)}$ denotes the index set of groups embraced by $G_i$, including $G_i$.

Since the crown scheduler features a predetermined execution order (cf. Section \ref{sec:intro}), the constraints previously controlling precedence relations and task start and completion times are now disposed of. The only remaining decision variables are the $x_{i,j,k}$, whose semantics is the same as for the group scheduler, and we still have $w_j = p_i$ given that $\tau_j$ is mapped to $G_i$. Regarding the ILP's constraints, solely \eqref{constr.b} is carried over from the group scheduler. Beyond that, two new constraints are adopted:
\begin{align}
&\forall j \quad \sum_{i:p_i>W_j,k} x_{i,j,k} = 0,\label{constr.crown1}\\
&\forall l \quad \sum_{i:l\in G_i,j,k} x_{i,j,k} \cdot t_j(p_i,f_k) \leq M\label{constr.crown2}.
\end{align}
Here, \eqref{constr.crown1} precludes a task from being mapped to a group whose size is larger than the task's maximum width. Constraint \eqref{constr.crown2} ensures that no core receives more work than it can handle until the deadline by requiring for each core $l$ that the accumulated runtime of tasks executed in any of the groups $l$ is a member of does not exceed $M$.

\section{Evaluation}\label{sec:eval}
We have conducted experiments with synthetic task sets, where $n \in \{ 4, 8, 16, 32 \}$. For each cardinality, we have created 10 task sets for a total of 40 task sets. The tasks' workloads $\lambda_j$ are randomly determined integers in $[1,100]$ and maximum widths were chosen randomly from $\{ 2^i \mid i = 0,\ldots,3 \}$, both based on a uniform distribution but under the restriction that $\lambda_j / W_j > 25$ when choosing $W_j$. Thus, no large tasks with low maximum width occur, which might call for loose deadlines to produce a feasible schedule in the first place. We have computed schedules for machines with 4 and 8 cores to cover the aspect of machine size. For any combination of task set size and machine size, four schedules per task set were determined via the four scheduling techniques presented in Section \ref{sec:constraints}: unrestricted scheduling, scheduling under allocation constraints, scheduling under allocation and group constraints, and crown scheduling.

All schedulers assume a generic core with power consumption modelled similar to 
ARM's big.LITTLE architecture 
\cite{kessler:2019}.
The parallel efficiency is computed as in \cite{MelotKKE14}:
\begin{gather*}
e_j(q) =
\begin{cases}
1 & \text{ for } q = 1,\\
1 - 0.3 \frac{q^2}{(W_j)^2} & \text{ for } 1 < q \leq W_j,\\
0.000001 & \text{ for } q > W_j,
\end{cases}
\end{gather*}
where $\tau_j$ is executed on $q$ cores, and the deadline is determined as in \cite{kessler:2019}:
\begin{gather*}
M =  d \cdot \frac{\sum_j \frac{\lambda_j}{p \cdot f_{max}} + 2 \sum_j \frac{\lambda_j}{p \cdot f_{min}}}{2},
\end{gather*}
where $d=0.8$ for $p=4$ and $d=1$ for $p=8$. These $d$ values were the lowest still yielding feasible solutions in all cases for the respective machine sizes. Here, $f_{min}$ and $f_{max}$ denote the machine's minimum and maximum operating frequencies, which in our case are \SI{0.6}{\GHz} and \SI{1.6}{\GHz}, cf. 
\cite{kessler:2019}.

For solving the ILPs, we have deployed the Gurobi 8.1.0 solver and the gurobipy module for Python. All schedules were computed on an AMD Ryzen 7 2700X with 8 cores and SMT. 
The ILP solver chooses itself how many of the up to 16 threads it uses.
The timeout was set to 5 minutes real (wall clock) time.

Aside from the schedules' total energy consumption as a measure of the schedules' quality the schedulers' execution time is of major interest in the present context. Generally speaking, solving an ILP is an expensive procedure, oftentimes requiring extensive computations. Table \ref{tab:timeouts} gives a first impression regarding the schedulers' resource consumption by presenting the number of timeouts reached for all combinations of scheduler, machine size, and task set size. As one can see, for small task sets of size 4, no timeouts have occurred. 
Large task sets of size 32 always lead to reaching the timeout. Differences between the four schedulers can only be observed for task set sizes of 8 and 16 and both machine sizes. As one would expect, the more constraints a scheduler is subject to, the fewer timeouts it encounters. The largest gap can be found between group and crown scheduler. When looking at task sets of size 16, the crown scheduler reaches the timeout in 1 of 20 cases, while all other schedulers never discover an optimal solution before the timeout occurs. On these grounds, one may surmise that the crown scheduler's predefined execution order -- its distinctive feature in our investigation --  substantially lowers the effort in the scheduling process.

\begin{table}
    \centering
    \caption{Number of timeouts for the schedulers under consideration and various combinations of task set size and machine size}
    \label{tab:timeouts}
    \begin{tabular}{cccccc}
        \toprule
         \# cores & \# tasks & unrestricted & allocpow2 & group & crown\\
         \midrule
        \multirow{5}{*}{4} & 4 & 0 & 0 & 0 & 0\\
         & 8 & 2 & 1 & 1 & 0\\
         & 16 & 10 & 10 & 10 & 0\\
         & 32 & 10 & 10 & 10 & 10\\
         \cmidrule{2-6}
         & total & 22 & 21 & 21 & 10\\
         \midrule
        \multirow{5}{*}{8} & 4 & 0 & 0 & 0 & 0\\
         & 8 & 3 & 1 & 1 & 0\\
         & 16 & 10 & 10 & 10 & 1\\
         & 32 & 10 & 10 & 10 & 10\\
         \cmidrule{2-6}
         & total & 23 & 21 & 21 & 11\\
         \midrule
         total & total & 45 & 42 & 42 & 21\\
         \bottomrule
    \end{tabular}
\end{table}

\begin{figure}
    \centering
    \includegraphics[width=0.42\textwidth]{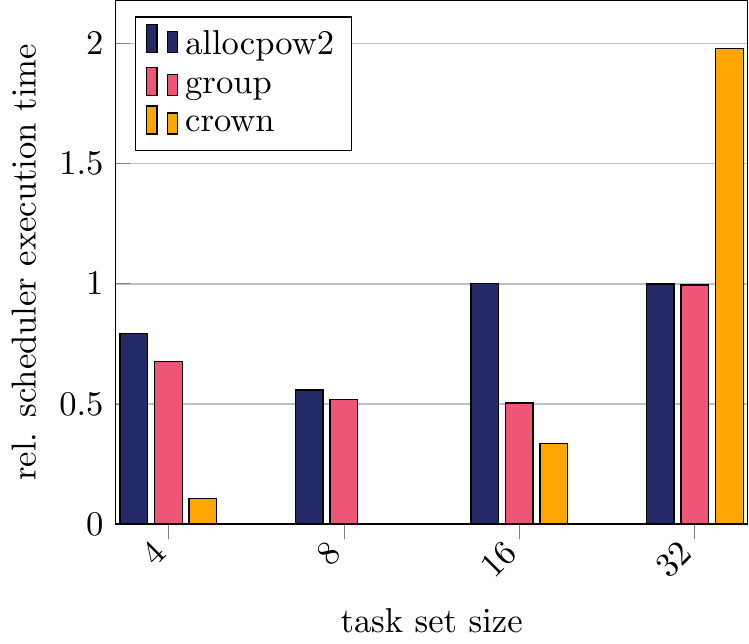}
    \hfill
    \includegraphics[width=0.42\textwidth]{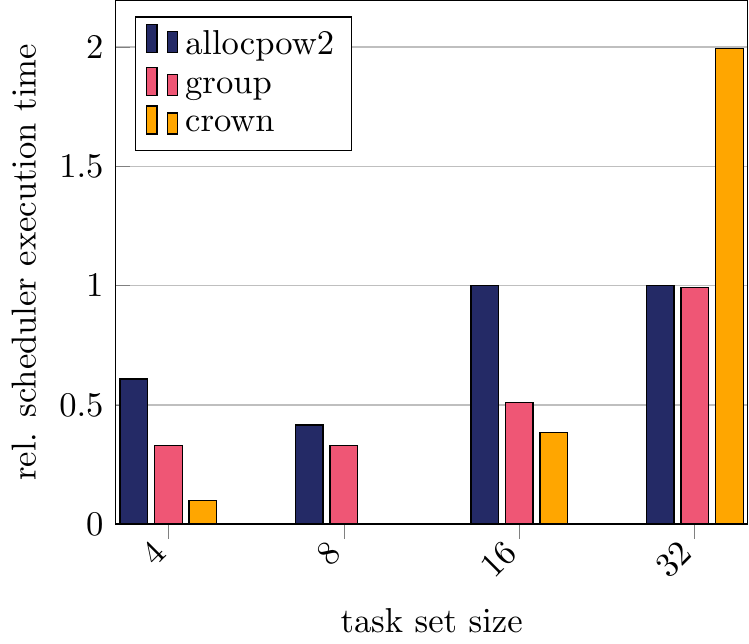}
    \caption{Scheduling times (CPU) for the schedulers under consideration grouped by task set size (values averaged over 10 task sets each), relative to the unrestricted scheduler. Left: 4-core machine, right: 8-core machine.}
    \label{fig:exectimesrel}
\end{figure}

To get a clearer picture, Table \ref{tab:exectimes} provides the average scheduling times (CPU times, i.\,e. sum of user and system times) and standard deviation for each combination of scheduler, machine size, and task set size.
Figure \ref{fig:exectimesrel} shows average scheduling time values for the constrained schedulers relative to the unrestricted scheduler.
We can see that the situation is similar for both machine sizes examined here. For very small task sets of size 4, all of the schedulers have produced solutions rapidly ($< \SI{1}{\s}$ of scheduler execution time). For the crown scheduler, this also applies to task sets of size 8 (the corresponding bar in fact is hardly noticeable), whereas the other schedulers' execution times are significantly longer. Here, restricting the allocation to powers of 2 halves scheduling time in relation to unrestricted scheduling, while adding the group constraints does not yield further gains. When looking at task sets of size 16, all schedulers but the crown scheduler constantly ran into the 5 minute wall clock timeout. Apparently, the unrestricted as well as the allocation-constrained scheduler were executed in 16 threads, while the group scheduler ran in 8 threads. This decision was made by the ILP solver. The crown scheduler not only makes do with roughly 35\% of the unrestricted scheduler's execution time, it also affords optimal solutions in all cases but one\footnote{One should note though that these solutions are optimal with regard to the crown scheduler's solution space, which is severely restricted in comparison to the unrestricted scheduler's. We will further consider solution quality below.}, cf. Table \ref{tab:optsols}. The largest gap in terms of resource consumption thus again opens up between the group and the crown scheduler. For large task sets of size 32, all schedulers have reached the timeout in any case. Interestingly, the crown scheduler was executed in 16 threads, while the other three schedulers ran in 8 threads (and therefore their CPU time is half the crown scheduler's). 
In most cases, standard deviation is fairly low indicating a roughly uniform scheduling time over all 10 task sets considered for a particular combination of machine size and task set size. For each scheduler, there is one task set size where standard deviation is high, suggesting that some task sets could be scheduled quickly and others took substantially longer, possibly even until timeout. Interestingly, the task set size in question is 16 for the crown scheduler and 8 for all other schedulers, leading to the conjecture that scheduling difficulty rises more slowly for the crown scheduler with increasing task set size.

\begin{table}[t]
    \centering
    \caption{Average scheduling times (CPU) and standard deviation values for the schedulers under consideration, for various combinations of task set size and machine size}
    \label{tab:exectimes}
    \begin{tabular}{crrrrrrrr}
        \toprule
        \# tasks & \multicolumn{2}{c}{unrestricted} & \multicolumn{2}{c}{allocpow2} & \multicolumn{2}{c}{group} & \multicolumn{2}{c}{crown}\\
        & time (\si{\second}) & st. dev. & time (\si{\second}) & st. dev. & time (\si{\second}) & st. dev. & time (\si{\second}) & st. dev.\\
        \midrule
        \multicolumn{9}{c}{4 cores}\\
        \midrule
        4 & 0.450 & 0.311 & 0.358 & 0.245 & 0.305 & 0.232 & 0.048 & 0.027\\
        8 & 1658.959 & 2105.047 & 927.608 & 1513.510 & 859.809 & 1542.652 & 0.444 & 0.269\\
        16 & 4757.157 & 30.308 & 4759.938 & 18.197 & 2397.118 & 0.524 & 1598.856 & 1280.000\\
        32 & 2393.429 & 1.212 & 2393.876 & 1.728 & 2381.849 & 2.181 & 4739.042 & 52.842\\
        \midrule
        \multicolumn{9}{c}{8 cores}\\
        \midrule
        4 & 0.478 & 0.350 & 0.291 & 0.256 & 0.158 & 0.125 & 0.047 & 0.019\\
        8 & 1716.667 & 2130.501 & 713.359 & 1486.013 & 567.248 & 1479.633 & 0.990 & 0.617\\
        16 & 4696.628 & 44.324 & 4699.877 & 30.329 & 2394.067 & 1.042 & 1797.960 & 2019.197\\
        32 & 2392.079 & 1.518 & 2392.394 & 0.701 & 2372.305 & 4.703 & 4771.125 & 2.076\\
        \bottomrule
    \end{tabular}
\end{table}

When it comes to the schedulers' performance in terms of solution quality, a first approach may be the number of optimal solutions each scheduler produces. From Table \ref{tab:optsols} one can gather that introducing the group constraints does not lead to an increase in optimal solutions discovered over the allocation-constrained scheduler. Both perform slightly better than the unrestricted scheduler though. The crown scheduler once again is far ahead of the other schedulers, mostly due to its strong performance for medium-sized task sets. One must keep in mind here that these figures reflect each scheduler's performance with regard to its own search space. Obviously, a smaller search space is beneficial when an optimal solution is to be found within a fixed period of time. 

\begin{table}
    \centering
    \caption{Number of optimal solutions for the schedulers under consideration and various combinations of task set size and machine size}
    \label{tab:optsols}
    \begin{tabular}{cccccc}
        \toprule
         \# cores & \# tasks & unrestricted & allocpow2 & group & crown\\
         \midrule
        \multirow{5}{*}{4} & 4 & 10 & 10 & 10 & 10\\
        & 8 & 8 & 9 & 9 & 10\\
        & 16 & 0 & 0 & 0 & 10\\
        & 32 & 0 & 0 & 0 & 0\\
        \cmidrule{2-6}
        & total & 18 & 19 & 19 & 30\\
        \midrule
        \multirow{5}{*}{8} & 4 & 10 & 10 & 10 & 10\\
        & 8 & 7 & 9 & 9 & 10\\
        & 16 & 0 & 0 & 0 & 9\\
        & 32 & 0 & 0 & 0 & 0\\
        & total & 17 & 19 & 19 & 29\\
        \midrule
        total & total & 35 & 38 & 38 & 59\\
         \bottomrule
    \end{tabular}
\end{table}

It is therefore of great interest to compare the energy consumption values for the schedules produced by the four schedulers. Table \ref{tab:energyvals} shows the respective values relative to the unrestricted scheduler's. For small task sets of 4 tasks, the constrained allocation leads to slightly higher energy consumption ($\approx 3\%$ on average). Further restrictions do not bring about yet another loss of solution quality. All schedules for the small task sets are optimal. Here, the unrestricted scheduler capitalizes on the more extensive search space. When task sets are larger, this benefit turns into a burden. Although the unrestricted scheduler's solution space comprises all the other schedulers' solution spaces, it does not manage to discover equally good solutions in due time. As one can see from Table \ref{tab:energyvals}, restricting the allocation does not change much in terms of energy consumption. Introducing additional group constraints in many cases does not have a massive impact, either. On the machine with 4 cores one can notice though that the deviation in both directions may be more pronounced: for the task sets with 16 tasks, the schedules' energy consumption is at 96\% of the unrestricted scheduler's on average, for the largest task sets with 32 tasks, it climbs to 114\%. Again, the most significant shift must be ascribed to the crown scheduler. For both machine sizes, the figures show a clear trend: the larger the task sets, the more energy is saved compared to the unrestricted scheduler. Since this observation does not apply to the group scheduler, one is lead to conjecture that the crown scheduler's predetermined execution order is the relevant factor enabling it to encounter higher quality solutions within a given time frame in relation to the other schedulers. Presumably, the execution order constraint considerably downsizes the search space without eliminating all the high quality solutions at the same time.

\begin{table}
    \centering
    \caption{Computed energy consumption values for the execution of the produced schedules, for the schedulers under consideration and for various combinations of task set size and machine size, relative to the unrestricted scheduler}
    \label{tab:energyvals}
    \begin{tabular}{ccc>{\bfseries}ccc>{\bfseries}ccc>{\bfseries}cc}
        \toprule
        \# cores & \# tasks & \multicolumn{3}{c}{allocpow2} & \multicolumn{3}{c}{group} & \multicolumn{3}{c}{crown}\\
        && best & avg. & worst & best & avg. & worst & best & avg. & worst\\
        \midrule
        \multirow{5}{*}{4} & 4 & 1.00 & 1.03 & 1.14 & 1.00 & 1.03 & 1.14 & 1.00 & 1.03 & 1.14\\
        & 8 & 1.00 & 1.00 & 1.01 & 1.00 & 1.00 & 1.01 & 1.00 & 1.00 & 1.01\\
        & 16 & 0.88 & 1.00 & 1.10 & 0.90 & 0.96 & 1.04 & 0.87 & 0.95 & 0.99\\
        & 32 & 0.88 & 0.99 & 1.11 & 0.88 & 1.14 & 1.84 & 0.83 & 0.89 & 0.97\\
        \cmidrule{2-11}
        & total & 0.88 & 1.00 & 1.14 & 0.88 & 1.04 & 1.84 & 0.83 & 0.97 & 1.14\\
        \midrule
        \multirow{5}{*}{8} & 4 & 1.00 & 1.03 & 1.15 & 1.00 & 1.03 & 1.15 & 1.00 & 1.03 & 1.15\\
        & 8 & 1.00 & 1.01 & 1.01 & 1.00 & 1.01 & 1.01 & 1.00 & 1.01 & 1.01\\
        & 16 & 0.99 & 1.00 & 1.03 & 0.98 & 0.99 & 1.00 & 0.97 & 0.98 & 0.99\\
        & 32 & 0.95 & 1.00 & 1.08 & 0.94 & 1.00 & 1.09 & 0.93 & 0.96 & 0.99\\
        \cmidrule{2-11}
        & total & 0.95 & 1.01 & 1.15 & 0.94 & 1.01 & 1.15 & 0.93 & 0.99 & 1.15\\
        \midrule
        total & total & 0.88 & 1.01 & 1.15 & 0.88 & 1.02 & 1.84 & 0.83 & 0.98 & 1.15\\
        \bottomrule
    \end{tabular}
\end{table}

All in all, in this section we have carved out that introducing allocation and group constraints yields similar solution quality when compared to an unrestricted scheduler, while scheduling time is significantly lower for small task sets. A further massive runtime decrease can be observed for the crown scheduler, as long as the timeout is not hit, which is constantly the case when task sets are large. Moreover, the crown scheduler's execution order constraints are likely to be credited with an improvement in solution quality, i.\,e. schedule energy consumption, over the other schedulers for large task sets. As we have seen, the gap broadens with increasing task set size. Only for very small task sets, the unrestricted scheduler delivers an uncontested performance. All these findings are largely independent of the machine size. Eventually, our investigation has revealed that solely constraining the allocation and potentially forming groups does not award the assets of the crown scheduling technique: a very low scheduling time when task sets are small, and a superior solution quality for larger task sets when scheduling time is limited. In nearly all scenarios, taking the additional step from group to crown scheduler thus pays off.

\section{Conclusions}\label{sec:concl}

We have presented a study on the evolution of scheduling time and energy efficiency of the resulting schedules when progressively constraining an unrestricted scheduler's search space, for sets of independent, non-preemptive, moldable tasks and parallel machines with discrete frequency levels.
Our studies indicate that constraining the tasks' execution order has most influence on both scheduler execution time and energy efficiency, given that scheduling time is constrained as well.
%
Thus, in most of the considered scenarios users are well-advised to deploy the crown scheduler, except for very small task sets, which is when the unrestricted scheduler can produce superior solutions without struggling with time constraints.

Future work will comprise the study of more fine-grained constraints. For example, one could first constrain assignments to consecutive processors, without being so strict as to only allow assignments within core groups.
Also, the order in which constraints are applied can be varied, for example assignment could be constrained before allocation.
Furthermore, evaluation shall be extended to task sets derived from real applications.

\subsection*{Acknowledgments}

We thank Christoph Kessler for many discussions and years -- past and future -- of fruitful and inspiring collaboration.

\bibliographystyle{splncs04}
\bibliography{biblio,references}

\begin{thebibliography}{10}
\providecommand{\url}[1]{\texttt{#1}}
\providecommand{\urlprefix}{URL }
\providecommand{\doi}[1]{https://doi.org/#1}

\bibitem{kessler:2019}
Kessler, C.W., Litzinger, S., Keller, J.: Adaptive crown scheduling for
  streaming tasks on many-core systems with discrete {DVFS}. In: Proc. 3rd
  International Workshop on Automatic Solutions for Parallel and Distributed
  Data Stream Processing (Auto-DaSP 2019) @ Euro-Par 2019 (2019)

\bibitem{MelotKEK19}
Melot, N., Kessler, C.W., Eitschberger, P., Keller, J.: Co-optimizing core
  allocation, mapping and {DVFS} in streaming programs with moldable tasks for
  energy efficient execution on manycore architectures. In: 19th International
  Conference on Application of Concurrency to System Design, {ACSD} 2019,
  Aachen, Germany, June 23-28, 2019. pp. 63--72 (2019)

\bibitem{MelotKKE14}
Melot, N., Ke{\ss}ler, C.W., Keller, J., Eitschberger, P.: Fast crown
  scheduling heuristics for energy-efficient mapping and scaling of moldable
  streaming tasks on manycore systems. ACM Trans. Archit. Code Optim.
  \textbf{11}(4),  62:1--62:24 (2014)

\bibitem{pruhs2008speed}
Pruhs, K., van Stee, R., Uthaisombut, P.: Speed scaling of tasks with
  precedence constraints. Theory of Computing Systems  \textbf{43}(1),  67--80
  (2008)

\bibitem{SandersSpeck2012}
Sanders, P., Speck, J.: Energy efficient frequency scaling and scheduling for
  malleable tasks. In: Euro-Par 2012 Parallel Processing. pp. 167--178.
  Springer LNCS (2012)

\bibitem{seo:2015}
Seo, W., Im, D., Choi, J., Huh, J.: Big or little: A study of mobile
  interactive applications on an asymmetric multi-core platform. In: 2015 IEEE
  International Symposium on Workload Characterization. pp. 1--11 (2015)

\bibitem{Turek1992}
Turek, J., Wolf, J.L., Yu, P.S.: Approximate algorithms scheduling
  parallelizable tasks. In: Proceedings of the Fourth Annual ACM Symposium on
  Parallel Algorithms and Architectures. pp. 323--332. SPAA '92, ACM, New York,
  NY, USA (1992)

\bibitem{xu12}
Xu, H., Kong, F., Deng, Q.: {E}nergy {M}inimizing for {P}arallel {R}eal-{T}ime
  {T}asks {B}ased on {L}evel-{P}acking. In: 18th Int.\ Conf.\ on Emb.\ and
  Real-Time Comput.\ Syst.\ and Appl.\ (RTCSA). pp. 98--103 (Aug 2012)

\bibitem{Ye2018}
Ye, D., Chen, D.Z., Zhang, G.: Online scheduling of moldable parallel tasks.
  Journal of Scheduling  \textbf{21}(6),  647--654 (Dec 2018)

\bibitem{Zahaf2017}
Zahaf, H.E., Benyamina, A.E.H., Olejnik, R., Lipari, G.: Energy-efficient
  scheduling for moldable real-time tasks on heterogeneous computing platforms.
  Journal of Systems Architecture  \textbf{74},  46--60 (2017)

\end{thebibliography}

\end{document}